\documentclass[%
 reprint,
nofootinbib,
 amsmath,amssymb,
 aps,
]{revtex4-2}

\usepackage{graphicx}
\usepackage{dcolumn}
\usepackage{bm}

\DeclareMathOperator{\Tr}{Tr}
\DeclareMathOperator{\tr}{tr}
\usepackage{xcolor}

\usepackage{hyperref}
\hypersetup{
    colorlinks = true,
    citecolor  = blue,
    linkcolor  = blue,
    urlcolor  = black
}

\begin{document}

\preprint{APS/123-QED}

\title{Zero-pairing and zero-temperature limits of finite temperature Hartree-Fock-Bogoliubov theory}

\author{T. Duguet}
 \email{thomas.duguet@cea.fr}
\affiliation{IRFU, CEA, Universit\'e Paris-Saclay, 91191 Gif-sur-Yvette, France}
\affiliation{KU Leuven, Instituut voor Kern- en Stralingsfysica, 3001 Leuven, Belgium}

\author{W. Ryssens}%
 \email{wouter.ryssens@yale.edu}
\affiliation{Center for Theoretical Physics, Sloane Physics Laboratory, 
                        Yale University, New Haven, CT 06520}%

\date{\today}

\begin{abstract}
\begin{description}
\item[Background] Recently, variational Hartree-Fock-Bogoliubov (HFB) mean-field equations were shown to possess a mathematically well-defined zero-pairing limit, independently of the closed- or open-shell character of the system under consideration. This limit is non-trivial for open-shell systems such that HFB theory does {\it not} reduce to the Hartree-Fock (HF) formalism in all cases.   
\item[Purpose] The present work extends this analysis to finite-temperature HFB (FTHFB) theory by investigating the behavior of this more general formalism in the combined zero-temperature and zero-pairing limits. 
\item[Methods] The zero-pairing and zero-temperature limits of the FTHFB statistical density operator constrained to carry an arbitrary (integer) number of particles A on average is worked out analytically and realized numerically using a two-nucleon interaction.
\item[Results] While the FTHFB density operator reduces to the projector corresponding to a pure HF Slater determinant for closed-shell nuclei,  the FTHFB formalism does not reduce to the HF theory in all cases in the zero-temperature and zero-pairing limits, i.e. for open-shell nuclei. However, the fact that a nucleus can be of open-shell character in these joint limits is necessarily the result of some symmetry restrictions. Whenever it is the case, the non-trivial description obtained for open-shell systems is shown to depend on the order with which both limits are taken, i.e. the two limits do not commute for these systems. When the zero-temperature limit is performed first, the FTHFB density operator is demoted to a projector corresponding to a pure state made out of a linear combination of a finite number of Slater determinants with different (even) numbers of particles. When the zero-pairing limit is performed first, the FTHFB density operator remains a statistical mixture of a finite number of Slater determinants with both even and odd particle numbers. While the entropy (pairing density) is zero in the first (second) case, it does not vanish in the second (first) case in spite of the temperature (pairing) tending towards zero. The difference between both limits can have striking consequences for the (thermal) expectation values of observables. For instance, the particle-number variance does not vanish in either case and has limiting values that differ by a factor of two in both cases.
\item[Conclusions] While in the textbook situation associated with closed-shell nuclei Hartree-Fock-Bogoliubov (finite-temperature Hartree-Fock) theory reduces to Hartree-Fock theory in the zero-pairing (zero-temperature) limit, the present analysis demonstrates that a non trivial and unexpected limit is obtained for this formalism in open-shell systems. This result sheds a new light on certain aspects of this otherwise very well-studied many-body formalism. 
\end{description}
\end{abstract}
\maketitle

\section{Introduction}

Hartree-Fock-Bogoliubov (HFB) theory~\cite{RiSc80} provides a variational mean-field approximation method to tackle pairing correlations in superfluid systems at the price of breaking $U(1)$ global gauge symmetry associated with particle-number conservation. When searching for the HFB solution, the particle number A is constrained on {\it average} to equate the physical value. In Ref.~\cite{Duguet:2020hdm}, the zero-pairing limit of the HFB formalism was investigated analytically and realized numerically. While in the textbook situation associated with closed-shell nuclei HFB theory reduces to Hartree-Fock (HF) theory, it was demonstrated that a non-trivial and unexpected solution is obtained in the limit for open-shell systems.

While many extensions of HFB theory exist to tackle low-lying excited states, the concept of discrete states loses its meaning at high excitation energy where the level density grows exponentially. In this regime, a statistical treatment of the system is more appropriate. The formalism extending HFB theory within the frame of statistical quantum mechanics is the so-called finite temperature HFB (FTHFB) formalism\footnote{Throughout the paper, HFB theory is referred to  as straight HFB to a priori distinguish it from the zero-temperature limit of FTHFB. Similarly, finite-temperature HF (FTHF) is considered to be possibly different from the zero-pairing limit of FTHFB.}. In this context, and with the goal to generalize the analysis of Ref.~\cite{Duguet:2020hdm}, it is of interest to investigate the FTHFB formalism in the zero-pairing and zero-temperature limits, both separately and jointly. While FTHFB theory trivially reduces to HF in the textbook case of closed-shell systems, the combined limits are presently shown to lead to a non-trivial and unexpected situation for open-shell systems. We note from the start that no system ends up being of open-shell character in the combined limits if the calculation is completely symmetry-unrestricted. The non-trivial situation we discuss in this paper is only encountered whenever one or more symmetry restrictions are imposed, which is very often the case in practical calculations. As a minimal symmetry restriction, we assume time-reversal invariance in all that follows. Eventually, further symmetry restrictions, e.g. rotational invariance, are considered. 

This paper is organized as follows. Section~\ref{sec:FTHFB} provides basic ingredients, including a short summary of the FTHFB formalism. While Sec.~\ref{sec:zerotemp} stipulates the definition of the zero-temperature and zero-pairing limits and how they are to be formally implemented, these limits are actually applied to the FTHFB formalism, first separately in Sec.~\ref{indiv_limits}, and then jointly in Sec.~\ref{comb_limits}. Next, Sec.~\ref{sec:numerical} displays the results of the numerical calculations illustrating the analytical conclusions reached in the previous sections. Eventually, Sec.~\ref{conclusions} provides the conclusions of the present work whereas a short appendix complements the paper.

\section{Many-body formalism}
\label{sec:FTHFB}

\subsection{Hamiltonian}

We start from a Hamiltonian whose second-quantized form is given by\footnote{Three-nucleon forces are presently omitted for simplicity given that none of the conclusions depend on their inclusion.}
\begin{align}
H &\equiv \sum_{ij} t_{ij}  c^{\dagger}_ic_j
             + \frac{1}{4} \sum_{ijkl} \bar{v}_{ijkl} c^{\dagger}_i c^{\dagger}_j c_l c_k \, . \label{hamiltonian}
\end{align}
In Eq.~\eqref{hamiltonian}, $H$ is expressed in terms of matrix elements of the one-body kinetic energy operator $\{t_{ij}\}$ and of the two-body interaction operator  $\{\overline{v}_{ikjl}\}$ in an arbitrary basis of the one-body space to which is associated a set of single-particle creation (annihilation) operators $\{c^{\dagger}_i\}$ ($\{c_i\}$).

\subsection{Finite-temperature framework}

The relevant thermodynamic potential to describe a nucleus at constant temperature T and chemical potential $\lambda$ is the grand potential~\cite{Goodman1981, BlaizotRipka, Schunck2019}
\begin{align}
\Omega &\equiv \text{E} - \text{T}\text{S} - \lambda \text{A} \, ,
\label{eq:grandpotential}
\end{align}
which is defined in terms of the average energy E, entropy S and particle number A\footnote{In actual applications, one Lagrange multiplier relates to constraining the neutron number N and one Lagrange multiplier is used to constrain the proton number Z. In our discussion A stands for either one of them.}. Both the energy and the average particle numbers are defined as thermal expectation values with respect to the (normalized) density operator\footnote{The use of a statistical density operator relates to an intrinsic lack of knowledge about the quantum state of the system. In the present case, the use of the grand potential corresponds to only knowing about the average energy and particle number of the system. In case a complete knowledge about the quantum state of the system can be assessed, the description is formulated in terms of a pure state.} ${\cal D}$ of the system\footnote{In Eqs.~\eqref{eq:def} and \eqref{eq:defS}, the uppercase $\Tr$ indicates a many-body trace over Fock space, while the lower case $\tr$ will be used to indicate traces over the one-body Hilbert space.}
\begin{subequations}
\label{eq:def}
\begin{align}
\text{E}_{{\cal D}}   &\equiv  \Tr \left[ {\cal D} H     \right] \, , \label{eq:defE} \\
\text{A}_{{\cal D}} &\equiv   \Tr \left[ {\cal D} A   \right] \, , \label{eq:defN}
\end{align}
\end{subequations}
while the entropy is calculated as
\begin{align}
\text{S}_{{\cal D}} &\equiv - \Tr \left[ {\cal D} \ln {\cal D} \right] \, . \label{eq:defS}
\end{align}

Requiring that the system is in thermal equilibrium is equivalent to minimizing the grand potential. Using the chemical potential as a Lagrange parameter to ensure the correct particle number on average, the minimization of $\Omega$ provides the formal solution
\begin{align}
    {\cal D} &= \frac{1}{{\cal Z}} e^{- \beta \left( H - \lambda A \right)} \, ,
    \label{eq:defdensityoperator}
\end{align}
where $\beta \equiv 1/\text{T}$ is the inverse temperature and where ${\cal Z}$ denotes the grand-canonical partition function
\begin{align}
    {\cal Z} = \Tr  \left[e^{- \beta \left( H - \lambda A \right)}\right]
\end{align}
ensuring the normalization of ${\cal D}$.

\subsection{Hartree-Fock-Bogoliubov approximation}

Given that the exact solution (Eq.~\eqref{eq:defdensityoperator}) is intractable for any realistic Hamiltonian, approximations must be formulated. The mean-field FTHFB approximation consists of using a trial density operator of the form\footnote{A partition function associated with $D$ can be defined but it is not equal to $Z$ given that the derivatives of the latter do not satisfy thermodynamic consistency relations~\cite{BlaizotRipka,Ryssens2020}.}
\begin{subequations}
\label{HFBdensityoperator}
\begin{align}
    D &\equiv \frac{1}{Z}e^{-\beta K} \, , \\
    Z &\equiv \Tr \left[e^{-\beta K}\right]\, .
\end{align}
\end{subequations}
where $K$ denotes a general (i.e. particle-number breaking) one-body operator
\begin{align}
        K &\equiv \frac{1}{2} \sum_{ij} \left[ k^{11}_{ij} 
         \left( c^{\dagger}_i c_j - c_j c^{\dagger}_i \right)
+  k^{20}_{ij} c^{\dagger}_i c^{\dagger}_j
+  k^{02}_{ij} c_j c_i \right] \nonumber \\
&\equiv  \frac{1}{2}\begin{pmatrix}
        c \\
        c^{\dagger}
        \end{pmatrix}^{\dagger}
        \mathcal{K}
        \begin{pmatrix}
        c \\
        c^{\dagger}
        \end{pmatrix}  \, , \label{onebodyK}
\end{align}
with 
\begin{align}
    \mathcal{K} &= 
    \begin{pmatrix}
        k^{11} & k^{20} \\
        - k^{02} & -k^{11\ast}
    \end{pmatrix} \, .
\end{align}
Using Wick's theorem for statistical mixtures~\cite{Gaudin1960}, the minimization of $\Omega$ with respect to the (independent) matrix elements defining $K$ can be shown to lead to~\cite{Goodman1981,BlaizotRipka}
\begin{align}
    \mathcal{K} &= \mathcal{H} \equiv
    \begin{pmatrix}
        h - \lambda & \Delta \\
        - \Delta^* & -h^* + \lambda
    \end{pmatrix} \, , \label{HFBsolD}
\end{align}
where the one-body fields $h$ and $\Delta$ making up the FTHFB Hamiltonian $\mathcal{H}$ are defined through
\begin{subequations}
\label{fields}
\begin{align}
    h_{ij}      &\equiv t_{ij} + \sum_{kl} \bar{v}_{iljk} \rho_{kl}  \equiv  t_{ij} + \Gamma_{ij} \, , \\
    \Delta_{ij} &\equiv \frac{1}{2} \sum_{kl} \bar{v}_{ijkl} \kappa_{kl}   \, .
\end{align}
\end{subequations}
In Eqs.~\eqref{fields}, the finite-temperature Hartree-Fock and Bogoliubov fields reads formally as in straight HFB theory but are expressed in terms of the finite-temperature normal ($\rho$) and anomalous ($\kappa$) one-body density matrices associated with $D$
\begin{subequations}
\label{onebodydensmatricesFTHFB}
\begin{align}
    \rho_{ij} &\equiv  \Tr \left( D c^{\dagger}_j c_i \right) \, , \label{onebodydensmatricesFTHFB_rho} \\
    \kappa_{ij} &\equiv  \Tr \Big( D c_j c_i \Big) \, . \label{onebodydensmatricesFTHFB_kappa} 
\end{align}
\end{subequations}

Finally, the FTHFB average particle number, total energy and particle-number variance can be written, via the application of Wick's theorem, as traces over the one-body space
\begin{subequations}
\label{eq:calc}
\begin{align}
\text{A}_{D}   &\equiv \Tr \left[ D A     \right]  \nonumber \\
               &=\tr \rho \, , \label{eq:calcN} \\              
\text{E}_{D} &\equiv \Tr \left[ D H     \right]  \nonumber \\
             &= \tr \left( t \rho \right)
               +\frac{1}{2} \tr \left( \Gamma \rho \right) 
              -\frac{1}{2} \tr \left(\Delta \kappa^{*} \right) \, , \label{eq:calcE} \\
\Delta \text{A}_{D}  &\equiv \Tr \left[ D A^2 \right] -  \Tr \left[D A\right]^2  \nonumber \\
               &=\tr \left[ \rho (1 - \rho) \right] + \tr \left[ \kappa^{\dagger} \kappa \right]  \, . \label{eq:calcDeltaN} 
\end{align}
\end{subequations}
In what follows, the last term in Eq.~\eqref{eq:calcE} will be denoted as the pairing or Bogoliubov energy  $E^{\text{B}}_{D}$. 

\subsection{Quasi-particle basis}

It is most convenient to formulate the FTHFB formalism in the quasi-particle basis diagonalizing $\mathcal{H}$ according to
\begin{align}
\mathcal{H}
\begin{pmatrix}
U_{k} \\
V_{k} 
\end{pmatrix}
&=
E_{k}
\begin{pmatrix}
U_{k} \\
V_{k} 
\end{pmatrix}
\label{eq:Hhfbdiag}\, ,
\end{align}
where the eigenvalues $\{E_k\}$ denotes the so-called FTHFB quasi-particle energies. The eigenvectors in Eq.~\eqref{eq:Hhfbdiag} define a set of quasi-particle creation and annihilation operators through the unitary Bogoliubov transformation~\cite{RiSc80}
\begin{subequations}
\begin{align}
\beta_{k} &\equiv \sum_i U^*_{ik} c_i + \sum_i V^*_{ik} c^\dagger_i\, , \\
\beta_{k}^\dagger &\equiv \sum_i U_{ik} c^\dagger_i + \sum_i V_{ik} c_i \, .
\end{align}
\end{subequations}
In matrix form, the transformation can be written as
\begin{equation}
\left(
\begin{array} {c}
\beta \\
\beta^{\dagger}
\end{array}
\right) = \mathcal{W}^{\dagger} \left(
\begin{array} {c}
c \\
c^{\dagger}
\end{array}
\right) \, , \label{bogotransfomatrixform}
\end{equation}
with the Bogoliubov matrix reading as
\begin{equation}
\mathcal{W} \equiv \left(
\begin{array} {cc}
U & V^{\ast} \\
V &  U^{\ast}
\end{array}
\right) \, . \label{bogomatrix}
\end{equation}
The unitarity of $\mathcal{W}$ ensures that the quasi-particle operators fulfill standard fermionic anticommutation rules.

Limiting the present study to time-reversal invariant systems, the FTHFB generalized density matrix is built from the eigenvectors of ${\cal H}$ with positive eigenvalues according to
\begin{align}
    \mathcal{R} &\equiv
    \begin{pmatrix}
    \rho & \kappa \\
    -\kappa^* & 1 - \rho^* 
    \end{pmatrix} = 
    \mathcal{W}
    \begin{pmatrix}
        f & 0 \\
        0 & 1 -f 
    \end{pmatrix}
    \mathcal{W}^{\dagger} \, ,
    \label{eq:genden}
\end{align}
where the matrix $f$ is diagonal and composed of the quasi-particle occupation factors~\cite{BlaizotRipka}
\begin{align}
    f_{kl} &\equiv  \Tr \left( D \beta^{\dagger}_l \beta_k \right) \nonumber \\
    &=\frac{1}{1 + e^{\beta E_k}} \, \delta_{kl} \nonumber \\
    &\equiv f_k  \, \delta_{kl} \, . \label{occfactor}
\end{align}
Contrary to the situation encountered in straight HFB theory, the FTHFB generalized density matrix is not idempotent, i.e. $\mathcal{R}^2\neq \mathcal{R}$. Given Eq.~\eqref{eq:genden}, the one-body density matrices in Eq.~\eqref{onebodydensmatricesFTHFB} are obtained according to~\cite{BlaizotRipka}
\begin{subequations}
\label{onebodydensmatricesFTHFB_bogo}
\begin{align}
    \rho_{ij} &\equiv \left(V^{\ast}(1-f)V^{T} \right)_{ij} + \left(UfU^{\dagger} \right)_{ij} \, , \label{onebodydensmatricesFTHFB_bogo_rho} \\
    \kappa_{ij} &\equiv  \left(V^{\ast}(1-f)U^{T} \right)_{ij} + \left(UfV^{\dagger} \right)_{ij} \, , \label{onebodydensmatricesFTHFB_bogo_kappa} 
\end{align}
\end{subequations}
in terms of which the average particle-number constraint is written as
\begin{align}
\text{A}_{D}   &=  \sum_{i} \rho_{ii}  = \text{A}\, . \label{partnumbconstr} 
\end{align}

Given ${\cal W}$, the even-number-parity Bogoliubov reference state $| \Phi \rangle$ is introduced as the vacuum of the quasi-particle operators, i.e. as the many-body state defined through $\beta_k | \Phi \rangle = 0$ for all $k$. This state breaks $U(1)$ global gauge symmetry associated with the conservation of particle number, i.e. it is typically not an eigenstate of the particle number operator $A$. Combining $| \Phi \rangle$ with the set of many-body states generated via the creation of an arbitrary (even or odd) number of quasi-particle excitations on top of it
\begin{equation}
| \Phi^{k_1 k_2\ldots} \rangle \equiv \beta^{\dagger}_{k_1} \, \beta^{\dagger}_{k_2} \, \ldots | \Phi \rangle \, , \label{qpexcitations}
\end{equation}
a complete basis of Fock space is obtained. 

Employing Eqs.~\eqref{eq:Hhfbdiag}-\eqref{bogotransfomatrixform}, the operator $K$ is easily re-expressed as
\begin{align}
    K &=  -\frac{1}{2} \sum_{k} E_k + \sum_k E_k \beta^{\dagger}_k \beta_k \, ,
    \label{rewrite_K}
\end{align}
where the sums run over positive eigenvalues of $\mathcal{H}$. Introducing
\begin{align}
    Z_{1} &\equiv Z \, e^{-\frac{\beta}{2} \sum_{k} E_k} = \Tr  \left[e^{-\beta \sum_k E_k \beta^{\dagger}_k \beta_k}\right] \, ,
\end{align}
the FTHFB density operator takes the simplified form
\begin{align}
    D &= \frac{1}{Z_{1}} e^{-\beta \sum_k E_k \beta^{\dagger}_k \beta_k} \, .
    \label{rewrite_FTHFB_densityoperator}
\end{align}
With Eq.~\eqref{rewrite_FTHFB_densityoperator} at hand, elementary commutation relations and the application of Baker-Campbell-Hausdorff’s identity allow one to prove that
\begin{subequations}
\label{commut_D}
\begin{align}
    D| \Phi \rangle &= \frac{1}{Z_{1}} | \Phi \rangle \, , \label{D_vac} \\
    D\beta^{\dagger}_k D^{-1} &= \xi_{k} \, \beta^{\dagger}_k \, , \label{commut_D_crea} \\
    D\beta_k D^{-1} &= \xi^{-1}_{k} \beta_k \, , \label{commut_D_anni}
\end{align}
\end{subequations}
where the statistical weight of a quasi-particle excitation is given by
\begin{align}
  \xi_{k} &\equiv e^{-\beta E_k}  \, . \label{stat_weight}
\end{align}
Using the completeness relation associated with the many-body basis of Fock space introduced in Eq.~\eqref{qpexcitations} and employing Eqs.~\eqref{D_vac}-\eqref{commut_D_crea} repeatedly, one obtains
\begin{align}
    D &= \frac{1}{Z_{1}}\Big(| \Phi \rangle \langle \Phi | \nonumber \\
    &\hspace{1cm} + \sum_{k_1} \xi_{k_1} | \Phi^{k_1} \rangle \langle \Phi^{k_1} | \nonumber \\
    &\hspace{1cm} + \frac{1}{2!} \sum_{k_1k_2} \xi_{k_1}\xi_{k_2} | \Phi^{k_1k_2} \rangle \langle \Phi^{k_1k_2} | \nonumber \\
    &\hspace{1cm} + \frac{1}{3!} \sum_{k_1k_2k_3} \xi_{k_1}\xi_{k_2}\xi_{k_3} | \Phi^{k_1k_2k_3} \rangle \langle \Phi^{k_1k_2k_3} | \nonumber \\
    &\hspace{1cm} + \ldots \Big) \, , \label{explicitD}
\end{align}
where 
\begin{align}
    Z_{1} &= \prod_k \left(1+ \xi_{k}\right) \, . \label{explicitZ}
\end{align}
One observes that, even for systems characterized by time-reversal invariance, the FTHFB density operator involves Bogoliubov states carrying both even and odd number-parity. 

Finally, the entropy can also be conveniently rewritten as
\begin{align}
\text{S}_{D}  = -\sum_k \left[f_k \ln f_k + (1-f_k) \ln (1 -f_k)\right] \, .
\label{eq:entropyHFB}
\end{align}

\subsection{Canonical basis}
\label{cano_basis}

As in straight HFB theory, the Bogoliubov vacuum can be most conveniently written in its canonical, i.e., BCS-like, form~\cite{RiSc80}
\begin{equation}
| \Phi \rangle \equiv \prod_{k>0} \left[u_k + v_k \, a^{\dagger}_k a^{\dagger}_{\bar{k}}\right] | 0 \rangle \, . \label{HFBstate}
\end{equation}
In Eq.~\eqref{HFBstate}, operators $\{ a^{\dagger}_k, a_k\}$ characterize the so-called canonical one-body basis in which pairs of conjugate states $(k,\bar{k})$ are singled out by the Bogoliubov transformation.  Conventionally, the two members of such a pair are distinguished as $k>0$ and $\bar{k}<0$, effectively splitting the basis into two halves. The coefficients $u_k=+u_{\bar{k}}$ and $v_k=-v_{\bar{k}}$ are BCS-like occupation numbers making up the canonical part of the full Bogoliubov transformation obtained through the Bloch-Messiah-Zumino decomposition~\cite{RiSc80} of the latter. The canonical Bogoliubov transformation is $2\times 2$ block diagonal and only couples conjugate single-particle states to generate conjugate quasi-particle operators according to
\begin{subequations}
\label{canoBogotransfo}
\begin{align}
\alpha^{\dagger}_{k} &=  u_{k} a^{\dagger}_{k} - v_k    a_{\bar{k}} \, , \\
\alpha^{\dagger}_{\bar{k}} &=  u_{k} a^{\dagger}_{\bar{k}} + v_k    a_k \, ,
\end{align}
\end{subequations}
whose hermitian conjugates annihilate $| \Phi \rangle$. The BCS-like occupation numbers can be chosen real,  satisfy the identity $u^2_k + v^2_k = 1$ and take the explicit form
\begin{align}
v_k^2 &\equiv \frac{1}{2} \left(1 - \frac{\epsilon_k - \lambda}{\sqrt{(\epsilon_k - \lambda)^2 + \Delta_k^2}}\right)  \, , \label{v2}
\end{align}
where $\epsilon_k \equiv h_{kk} = h_{\bar{k}\bar{k}}$ and $\Delta_k \equiv \Delta_{k\bar{k}} = - \Delta_{\bar{k}k}$.

\section{Zero temperature and pairing limits}
\label{sec:zerotemp}

The objective of the present work is to study the FTHFB formalism in the combined limits of vanishing pairing and temperature, while keeping the average particle number fixed to an integer value A. The present analysis extends the detailed study of the zero-pairing limit within the HFB formalism given in Ref.~\cite{Duguet:2020hdm}. 

While each individual limit of the FTHFB formalism is straightforward, the main outcome of the present study is the fact that both limits do not commute in general when taken together: a well-defined solution exists for any A but the nature of that solution depends on the order with which the two limits are performed. Four cases are to be distinguished
\begin{enumerate}
\item Individual limits
\begin{enumerate}
\item $\text{T}\rightarrow 0 \, , \Delta\neq 0$ 

Zero-temperature limit at finite pairing.
\item $\Delta\rightarrow 0  \,  , \text{T}\neq 0$ 

Zero-pairing limit at finite temperature.
\end{enumerate}
\item Combined limits
\begin{enumerate}
\item $\text{T}\rightarrow 0 \, \& \, \Delta \rightarrow 0$ 

Zero-temperature limit followed by the zero-pairing limit. 
\item $\Delta\rightarrow 0 \, \& \, \text{T}\rightarrow 0$

Zero-pairing limit followed by the zero-temperature limit.
\end{enumerate}
\end{enumerate}
such that case 2.a (2.b) is nothing but case 1.a (1.b) on top of which the zero-pairing (zero-temperature) limit is further performed. The necessity to distinguish cases 2.a and 2.b  relates to the fact that the two limits do not commute.

\subsection{Implementation}

Taking the zero-temperature and zero-pairing limits of the FTHFB formalism corresponds to operating specific mathematical limits under the condition that the constraint on the average particle-number (Eq.~\eqref{partnumbconstr}) is satisfied. Let us now briefly specify these mathematical operations in the two cases of interest before applying them, first separately, and then sequentially.

\subsubsection{Zero-temperature limit}

The zero-temperature limit is straightforwardly realized by applying the operation $\text{T}\rightarrow 0$ or $\beta \rightarrow \infty$  in Eq.~\eqref{HFBdensityoperator}, under the condition that the constraint on the average particle-number (Eq.~\eqref{partnumbconstr}) is satisfied. Employing the basis of Fock space built from the Bogoliubov reference state, the procedure translates into taking this mathematical limit in Eqs.~\eqref{eq:genden}-\eqref{onebodydensmatricesFTHFB_bogo_kappa} while satisfying Eq.~\eqref{partnumbconstr}.

\subsubsection{Zero-pairing limit}

The zero-pairing limit is materialized by scaling the Bogoliubov field $\Delta$ down to zero in the HFB Hamiltonian ${\cal H}$ under the condition that Eq.~\eqref{partnumbconstr} is satisfied, i.e.
\begin{equation}
\Delta_{ij} \rightarrow 0 \,\,\,\, \forall (i,j) \,\,\, \text{subject to} \,\,\, \text{A}_{D}  = \text{A} \label{limit} \, .
\end{equation}

In practice, the limit is achieved by adding a constraining term to the grand potential $\Omega$ such that the operator used for the minimization becomes the Routhian~\cite{Duguet:2020hdm}
\begin{align}
\Omega(\delta)_{D} = \Omega_{D}   - \frac{1}{2}(1 -\delta)(\Delta_C)_{D} \, .    \label{constrainedOmega}
\end{align}
where the Hermitian operator $\Delta_C$ is defined through
\begin{align}
    \Delta_C \equiv \frac{1}{2} \sum_{ij} \Delta_{ij} c^{\dagger}_i c^{\dagger}_j  + \frac{1}{2} \sum_{ij} \Delta^*_{ij} c_j c_i \, . 
\end{align}
The thermal trace of $\Delta_C$ is exactly \emph{twice} the pairing energy of Eq.~\eqref{eq:calcE}. The FTHFB density operator obtained through the minimization of $\Omega(\delta)_{D}$ is formally the same as before except that the HFB Hamiltonian in Eq.~\eqref{HFBsolD} must be replaced by
\begin{align}
    \mathcal{H}(\delta) &= 
    \begin{pmatrix}
    h - \lambda & \delta \Delta \\
    - \delta \Delta^* & -h^* + \lambda
    \end{pmatrix} \, . \label{constrainedH}
\end{align}
in which the original pairing field is now multiplied with the prefactor $\delta$. While the unconstrained formalism is recovered for $\delta = 1$, the zero-pairing limit corresponds to taking $\delta\rightarrow 0$ in Eq.~\eqref{constrainedH}, i.e. to fully subtracting the pairing energy from the grand potential such that the pairing field is zero in $\mathcal{H}(\delta)$.

\subsection{Naive filling}
\label{naivefilling}

The zero-temperature and zero-pairing limits of the FTHFB formalism rely on the {\it naive filling} of canonical shells\footnote{A nuclear ``shell" presently stands for a collection of degenerate single-particle levels, independently of the symmetry responsible for their actual degree of degeneracy. While it naturally encompasses the particular case of ``spherical" shells, this definition is more general and thus also valid whenever the degree of symmetry is lower.} characterizing the system of interest when reaching these limits. 

The naive filling corresponds to occupying canonical single-particle states characterized by the A lowest energies $\epsilon_k$. Doing so, one partitions the A nucleons in such a way that $a_v$ nucleons sit in the so-called {\it valence}, i.e., last occupied, shell characterized by energy $\epsilon_v$ and degeneracy $d_v$ (i.e. $p_v\equiv d_v/2$ pairs of conjugate states). The naive occupation of each canonical state belonging to the valence shell
\begin{equation}
o_v \equiv  \frac{a_v}{d_v} \, , \label{meanoccvalenceshell}
\end{equation}
ranges between 0 and 1, i.e., $0<o_v \leq 1$. 

It is natural to distinguish three categories of canonical single-particle states, i.e. states characterized by
\begin{enumerate}
\item $\epsilon_h - \lambda <0$, casually denoted as ``hole states",
\item $\epsilon_v - \lambda =0$, casually denoted as ``valence states",
\item $\epsilon_p - \lambda >0$, casually denoted as ``particle states",
\end{enumerate}
when reaching the limits, such that valence states can only concern one shell. 

Two different classes of nuclei emerge in this context, i.e., a nucleus is either of closed-shell character when $o_v=1$ or of open-shell character whenever $0<o_v < 1$. That a given nucleus belongs to one category or the other can only be inferred \textit{a posteriori} and depends on the symmetries, and thus on the degeneracies, characterizing the spectrum $\{\epsilon_k\}$ in the limits. For example, a nucleus qualifying as a {\it spherical} open-shell system whenever spherical symmetry is enforced can turn into a {\it deformed} closed-shell system whenever SU(2) symmetry is allowed to break\footnote{In symmetry-fully-unrestricted calculations, a {\it spherical} even-even open-shell system typically has energetic advantage to lift the $(2j+1)$-fold degeneracy associated with spherical symmetry such as to reach a {\it deformed} closed-shell configuration in the combined zero-pairing and zero-temperature limits. Consequently, the encounter of open-shell systems typically relates to specific symmetry constraints built into the numerical code/calculation.}. In the present context, the notions of closed- and open-shell systems are not restricted to a specific symmetry, e.g. a closed-shell nucleus can be of spherical or deformed character.

\section{Individual limits}
\label{indiv_limits}

Let us first consider the two limits separately.

\subsection{$\text{T} \rightarrow 0,  \Delta \neq 0$}
\label{case_1a}

The first case of interest corresponds to taking the zero-temperature limit whenever pairing, i.e. the pairing field in the HFB Hamiltonian, is non-zero. The hypothesis that pairing does not vanish implies that all quasi-particle energies are {\it strictly} positive, $E_k>0, \,\, \forall k$. As a result, Eqs.~\eqref{occfactor} and~\eqref{stat_weight} stipulate that
\begin{subequations}
\begin{align}
\lim_{\underset{\text{A}_{D}  = \text{A}}{\text{T}\rightarrow 0/\Delta\neq 0}}    f_{k} &= 0  \,\,\,\, \forall k \, , \\
\lim_{\underset{\text{A}_{D}  = \text{A}}{\text{T}\rightarrow 0/\Delta\neq 0}}   \xi_k &= 0   \,\,\,\, \forall k \, ,
\end{align}
\end{subequations}
such that Eqs.~\eqref{eq:genden}, \eqref{explicitD} and~\eqref{explicitZ} deliver
\begin{subequations}
\begin{align}
 \lim_{\underset{\text{A}_{D}  = \text{A}}{\text{T}\rightarrow 0, \Delta\neq 0}}     \mathcal{R} &= 
    \mathcal{W}
    \begin{pmatrix}
        0 & 0 \\
        0 & 1
    \end{pmatrix}
    \mathcal{W}^{\dagger} \, , \\
\lim_{\underset{\text{A}_{D}  = \text{A}}{\text{T}\rightarrow 0/\Delta\neq 0}}   D &= | \Phi \rangle \langle \Phi |  \, .
\end{align}
\end{subequations}
The density operator describing the system reduces to the projector associated with the even-number-parity HFB vacuum characterized by $\langle \Phi | A | \Phi \rangle = \text{A}$, i.e. a pure state, such that the FTHFB formalism trivially reduces to straight HFB theory in which the generalized density matrix is idempotent. Correspondingly, one has
\begin{align}
\lim_{\underset{\text{A}_{D}  = \text{A}}{\text{T}\rightarrow 0, \Delta\neq 0}} \text{S}_{D} &= 0  \, .
\end{align}
In this context, the one-body density matrices take the following simple form in the canonical basis
\begin{subequations}
\label{1bodydensmatlimits0}
\begin{align}
\lim_{\underset{\text{A}_{D}  = \text{A}}{\text{T}\rightarrow 0, \Delta\neq 0}} \rho_{kk'} & =  v^{2}_k \, \delta_{kk'} \, , \label{limith0} \\
\lim_{\underset{\text{A}_{D}  = \text{A}}{\text{T}\rightarrow 0, \Delta\neq 0}} \kappa_{kk'} & =  u_k v_k \, \delta_{\bar{k}k'}   \, .
\end{align}
\end{subequations}
Furthermore, the idempotency of the generalized density matrix can be used to simplify the expression of the particle-number variance such that Eqs.~\eqref{eq:calcN},\eqref{eq:calcE} and \eqref{eq:calcDeltaN} become
\begin{subequations}
\label{eq:calcnew}
\begin{align}
\lim_{\underset{\text{A}_{D}  = \text{A}}{\text{T}\rightarrow 0,\Delta\neq 0}} \text{A}_{D}   &=\tr \rho \, , \label{eq:calcNZT} \\              
\lim_{\underset{\text{A}_{D}  = \text{A}}{\text{T}\rightarrow 0,\Delta\neq 0}} \text{E}_{D} &= \tr \left(t \rho \right)
               +\frac{1}{2} \tr \left( \Gamma \rho \right) 
              -\frac{1}{2} \tr \left(\Delta \kappa^{*} \right) \, , \label{eq:calcEZT} \\
\lim_{\underset{\text{A}_{D}  = \text{A}}{\text{T}\rightarrow 0,\Delta\neq 0}} \Delta \text{A}_{D} &= 2\tr \left[ \rho (1 - \rho) \right]\, . \label{eq:dispersionZT}
\end{align}
\end{subequations}

\subsection{$\Delta\rightarrow 0,  \text{T}\neq 0$}
\label{case_1b}

Let us now consider the zero-pairing limit at finite temperature. Taking the limit $\delta \rightarrow 0$ in Eq.~\eqref{constrainedH}, the operator $K$ (Eq.~\eqref{onebodyK}) becomes in the basis diagonalizing $h$ and $\rho$
\begin{align}
\lim_{\underset{\text{A}_{D}  = \text{A}}{\Delta\rightarrow 0,\text{T}\neq 0}}  K =  -\frac{1}{2} \sum_{k}  (\epsilon_{k} - \lambda) + \sum_{k}  (\epsilon_{k} - \lambda) a^{\dagger}_{k} a_{k} \, , \label{limitKzeropairing}
\end{align}
where the sums run over all eigenstates of $h$. This expression can easily be related to Eq.~\eqref{rewrite_K} through the behavior of the quasi-particle energies
\begin{align}
\lim_{\underset{\text{A}_{D}  = \text{A}}{\Delta\rightarrow 0,\text{T}\neq 0}}  E_k = |\epsilon_k - \lambda| \,\, \,\, \forall k \, , \label{limitQPE}
\end{align}
such that
\begin{align}
\lim_{\underset{\text{A}_{D}  = \text{A}}{\Delta\rightarrow 0,\text{T}\neq 0}}   f_{k} &= \frac{1}{1 + e^{\beta |\epsilon_k - \lambda|}} \,\,\,\, \forall k \, . \label{limitocc_zeropairingTdiff0_f} 
\end{align}
Given Eq.~\eqref{limitKzeropairing}, and redefining the normalization factor as
\begin{align}
    Z_{2} &\equiv  \Tr  \left[e^{-\beta \sum_{k}  (\epsilon_{k} - \lambda) a^{\dagger}_{k} a_{k}}\right] \, , \label{Z2}
\end{align}
the FTHFB density operator becomes
\begin{align}
\lim_{\underset{\text{A}_{D}  = \text{A}}{\Delta\rightarrow 0,\text{T}\neq 0}}      D &= \frac{1}{Z_{2}} e^{-\beta \sum_{k}  (\epsilon_{k} - \lambda) a^{\dagger}_{k} a_{k}} \, .
    \label{FTHFB_densityoperator_zeropairing}
\end{align}
Equations~\eqref{Z2}-\eqref{FTHFB_densityoperator_zeropairing} define nothing but the density operator at play in the finite-temperature Hartree-Fock (FTHF) formalism~\cite{BlaizotRipka} such that the FTHFB formalism strictly reduces to it in the zero-pairing limit. Given the form of the density operator, one-body density matrices are straightforwardly shown to satisfy
\begin{subequations}
\label{limitdensmatrices_zero_pairing}
\begin{align}
\lim_{\underset{\text{A}_{D}  = \text{A}}{\Delta\rightarrow 0,\text{T}\neq 0}} \rho_{kk'} &=  \frac{1}{1+ e^{\beta(\epsilon_k-\lambda)}} \delta_{kk'} \equiv f^{\text{HF}}_{k} \delta_{kk'} \, , \label{limitdensmatrices_zero_pairing1} \\
\lim_{\underset{\text{A}_{D}  = \text{A}}{\Delta\rightarrow 0,\text{T}\neq 0}} \kappa_{kk'} &=  0 \, , \label{limitdensmatrices_zero_pairing2}
\end{align}
\end{subequations}
such that, as expected, the anomalous density matrix is identically zero whereas the normal density matrix is diagonal in the HF single-particle basis.

It is interesting to look more carefully into how the FTHF formalism is obtained from the FTHFB one in all cases. In particular, noticing from Eq.~\eqref{limitQPE} that quasi-particle energies are not necessarily strictly positive in the zero-pairing limit, the fact that a valence shell characterized by $|\epsilon_v - \lambda| = 0$ may emerge must be contemplated with care. 

Ignoring this possibility at first, the textbook situation is encountered where nuclear shells define either hole ($\epsilon_h - \lambda<0$) or particle ($\epsilon_p - \lambda>0$) states.  In this case, the most convenient basis of Fock space to expand the density operator is given by the closed-shell reference Slater determinant 
\begin{align}
| \bar{\Phi} \rangle \equiv \prod_{h=1}^{A/2} a^{\dagger}_h a^{\dagger}_{\bar{h}}| 0 \rangle \, , \label{HFslatervacuum}
\end{align}
which is an eigenstate of $A$ with eigenvalue A, along with the complete set of Slater determinants obtained via $n$-particle,$m$-hole excitations on top of it
\begin{equation}
| \bar{\Phi}^{p_1 \ldots p_n}_{h_1 \ldots h_m} \rangle \equiv a^{\dagger}_{p_1} \ldots a^{\dagger}_{p_n} a_{h_m} \ldots a_{h_1} | \bar{\Phi} \rangle \, , \label{phexcitations}
\end{equation}
where the maximum number of annihilation operators is A whereas the number of creation operators is unrestricted. Given Eqs.~\eqref{Z2}-\eqref{FTHFB_densityoperator_zeropairing}, Eq.~\eqref{commut_D} is presently replaced by 
\begin{subequations}
\label{commut_D_zeropairing}
\begin{align}
\lim_{\underset{\text{A}_{D}  = \text{A}}{\Delta\rightarrow 0,\text{T}\neq 0}}    D|  \bar{\Phi} \rangle &= \frac{1}{Z_{2}} e^{-\beta \sum_{h}  (\epsilon_{h} - \lambda)} |  \bar{\Phi} \rangle \, , \label{D_vac_zeropairing} \\
\lim_{\underset{\text{A}_{D}  = \text{A}}{\Delta\rightarrow 0,\text{T}\neq 0}}    D a^{\dagger}_{k} D^{-1} &= \zeta_{k} \, a^{\dagger}_{k} \, , \label{commut_D_crea_zeropairing} \\
\lim_{\underset{\text{A}_{D}  = \text{A}}{\Delta\rightarrow 0,\text{T}\neq 0}}    D a_{k} D^{-1} &= \zeta^{-1}_{k}  a_{k}  \, , \label{commut_D_anni_zeropairing}
\end{align}
\end{subequations}
where the statistical weight of a particle creation is defined as
\begin{align}
  \zeta_{k} &\equiv e^{-\beta (\epsilon_{k} - \lambda)}  \, . \label{stat_weight_zeropairing}
\end{align}
Using the completeness relation of the above basis of Fock space and utilizing Eq.~\eqref{commut_D_zeropairing} repeatedly, one obtains
\begin{align}
   \lim_{\underset{\text{A}_{D}  = \text{A}}{\Delta\rightarrow 0,\text{T}\neq 0}}    D &= \frac{1}{Z_{2}} e^{-\beta \sum_{h}  (\epsilon_{h} - \lambda)} \Big(| \bar{\Phi} \rangle \langle \bar{\Phi} | \nonumber \\
    &\hspace{1cm} + \sum_{h_1} \zeta^{-1}_{h_1}  | \bar{\Phi}_{h_1} \rangle \langle \bar{\Phi}_{h_1} | \nonumber \\
    &\hspace{1cm} + \sum_{p_1} \zeta_{p_1}  | \bar{\Phi}^{p_1} \rangle \langle \bar{\Phi}^{p_1} | \nonumber \\
    &\hspace{1cm} + \sum_{h_1p_1} \zeta^{-1}_{h_1}\zeta_{p_1}   | \bar{\Phi}^{p_1}_{h_1} \rangle \langle \bar{\Phi}^{p_1}_{h_1} | \nonumber \\
    &\hspace{1cm} + \frac{1}{2!} \sum_{h_1h_2} \zeta^{-1}_{h_1}\zeta^{-1}_{h_2}   | \bar{\Phi}_{h_1h_2} \rangle \langle \bar{\Phi}_{h_1h_2} | \nonumber \\
    &\hspace{1cm} + \frac{1}{2!} \sum_{p_1p_2} \zeta_{p_1}\zeta_{p_2}   | \bar{\Phi}^{p_1p_2} \rangle \langle \bar{\Phi}^{p_1p_2} | \nonumber \\
    &\hspace{1cm} + \ldots \Big) \, , \label{explicitD_zeropairing}
\end{align}
with 
\begin{align}
    Z_{2} &= \prod_{k} (1+\zeta_{k}) \, . \label{explicitZ_zeropairing}
\end{align}

Let us now turn to the case where a valence shell characterized by $|\epsilon_v - \lambda| = 0$ emerges at $\text{T}\neq 0$ in the zero-pairing limit. This valence shell gathers $p_v=d_v/2$ pairs of conjugate states generically denoted as $(v,\bar{v})$ and specified as $(v_1,v_{\bar{1}}), \ldots, (v_{p_v},v_{{\bar{p}_v}})$. The mean occupation of each valence state is  $f_{v}=1/2$ (Eq.~\eqref{limitdensmatrices_zero_pairing1}) whereas the statistical weight associated with the creation of a valence particle is given by $\zeta_{v}=1$ (Eq.~\eqref{stat_weight_zeropairing}). 

In order to expand the density operator on a basis of Fock space, one first observes that the Bogoliubov reference state becomes in the zero-pairing limit
\begin{align}
\lim_{\underset{\text{A}_{D}  = \text{A}}{\Delta\rightarrow 0,\text{T}\neq 0}} | \Phi \rangle &=  \prod_{k=1}^{p_v} (u_v + v_v \, a^{\dagger}_{v_{k}} a^{\dagger}_{v_{\bar{k}}})| \tilde{\Phi} \rangle \, , \label{limitreferencestate}
\end{align}
where $| \tilde{\Phi} \rangle$ denotes the closed-shell Slater determinant built by occupying the $\text{A}-a_v$ hole states
\begin{align}
| \tilde{\Phi} \rangle & \equiv \prod_{h=1}^{(A-a_v)/2} a^{\dagger}_h a^{\dagger}_{\bar{h}}| 0 \rangle \, . \label{coreSD}
\end{align}
In agreement with Eq.~\eqref{qpexcitations}, a complete basis of Fock space is obtained by creating arbitrary numbers of quasi-particles associated with the set of operators
\begin{subequations}
\label{qpexcitation_limit}
\begin{align}
\lim_{\underset{\text{A}_{D}  = \text{A}}{\Delta\rightarrow 0,\text{T}\neq 0}} \alpha^{\dagger}_{h} &=  - a_{\bar{h}} \, , \label{qpexcitation_limit_h} \\
\lim_{\underset{\text{A}_{D}  = \text{A}}{\Delta\rightarrow 0,\text{T}\neq 0}} \alpha^{\dagger}_{v} &=  u_v  a^{\dagger}_{v} - v_v a_{\bar{v}} \, , \label{qpexcitation_limit_v} \\
\lim_{\underset{\text{A}_{D}  = \text{A}}{\Delta\rightarrow 0,\text{T}\neq 0}}\alpha^{\dagger}_{p} &=  a^{\dagger}_{p}  \, , \label{qpexcitation_limit_p}
\end{align}
\end{subequations}
where $v_v$ ($u_v=\sqrt{1-v_v^2}$) is an arbitrary number between 0 and 1. It is in fact convenient, and more natural in the context of FTHF, to choose the closed-shell Slater determinant $| \tilde{\Phi} \rangle$ carrying $\text{A}-a_v$ particles as a reference state and to generate the basis of Fock space through $n$-particle,$m$-hole excitations of it. This choice corresponds to setting $u_v=1$ and $v_v=0$ in Eqs.~\eqref{limitreferencestate}-\eqref{qpexcitation_limit}. Doing so, the expansion of the density operator is fully consistent with Eqs.~\eqref{explicitD_zeropairing}-\eqref{explicitZ_zeropairing} if one considers the $d_v$ valence states as particle states\footnote{One could have equally chosen the closed-shell Slater determinant carrying $\text{A}-a_v+d_v$ particles as a reference state. This option would correspond to choosing $u_v=0$ and $v_v=1$ in Eqs.~\eqref{limitreferencestate}-\eqref{qpexcitation_limit}. With such a choice, the expansion of the density operator would still be consistent with Eqs.~\eqref{explicitD_zeropairing}-\eqref{explicitZ_zeropairing} but at the price of considering the $d_v$ valence states as hole states with statistical weight $\zeta^{-1}_{v}=1$.} with statistical weight $\zeta_{v}=1$.

Based on all the above, the observables of interest read in the zero-pairing limit as
\begin{subequations}
\label{eq:calc_limdelta}
\begin{align}
\lim_{\underset{\text{A}_{D}  = \text{A}}{\Delta\rightarrow 0,\text{T}\neq 0}} \text{A}_{D}   &=\tr \rho \label{eq:calcNlimdelta} \\
&= \sum_k f^{\text{HF}}_{k} \, , \nonumber \\              
\lim_{\underset{\text{A}_{D}  = \text{A}}{\Delta\rightarrow 0,\text{T}\neq 0}} \text{E}_{D} &= \tr \left(t \rho \right)
               +\frac{1}{2} \tr \left( \Gamma \rho \right) \label{eq:calcElimdelta} \\
&= \sum_k t_{kk} f^{\text{HF}}_{k}  
               +\frac{1}{2} \sum_{kk'}  \bar{v}_{kk'kk'} f^{\text{HF}}_{k}  f^{\text{HF}}_{k'}    \nonumber \, ,  \\
\lim_{\underset{\text{A}_{D}  = \text{A}}{\Delta\rightarrow 0,\text{T}\neq 0}}  \Delta \text{A}_{D}  &=\tr \left[ \rho (1 - \rho) \right]  \label{eq:calcDeltaNlimdelta}  \\
&= \sum_k f^{\text{HF}}_{k}(1-f^{\text{HF}}_{k})\, ,  \nonumber
\end{align}
\end{subequations}
where one observes that the pairing contribution to the energy has disappeared and that the formal expression of the particle-number variance is half of the expression found in the zero-temperature limit (Eq.~\eqref{eq:dispersionZT}). The entropy is furthermore given by
\begin{align}
\text{S}_{D}  = -\sum_k \left[f^{\text{HF}}_{k} \ln f^{\text{HF}}_{k} + (1-f^{\text{HF}}_{k}) \ln (1 -f^{\text{HF}}_{k})\right] \, .
\end{align}

All the above demonstrates that the FTHFB formalism strictly reduces {\it in all systems}, i.e. independently of A, to FTHF in the zero-pairing limit.

\section{Combined limits}
\label{comb_limits}

Now that the individual limits 1.a and 1.b have been worked out, the goal is to combine them to study cases 2.a and 2.b. 

\subsection{$\text{T}\rightarrow 0 \, \& \, \Delta \rightarrow 0$}
\label{case_2a}

Based on the result of Sec.~\ref{case_1a} above, this case boils down to taking the zero-pairing limit of the straight HFB formalism. This situation was discussed at length in Ref.~\cite{Duguet:2020hdm} and is only briefly summarized below.

\subsubsection{Closed-shell system} 

In a closed-shell system where canonical shells strictly separate into A hole ($\epsilon_h - \lambda<0$) states and the remaining particle ($\epsilon_p - \lambda>0$) states, Eq.~\eqref{1bodydensmatlimits0} becomes
\begin{subequations}
\label{1bodydensmatlimits}
\begin{align}
\lim_{\underset{\langle A  \rangle  = \text{A}}{\text{T}\rightarrow 0 \, \& \, \Delta \rightarrow 0}} \rho_{kk'} & = \lim_{\underset{\langle A  \rangle  = \text{A}}{\text{T}\rightarrow 0 \, \& \, \Delta \rightarrow 0}} v^{2}_k \, \delta_{kk'} \nonumber \\
&= \Theta(\lambda -\epsilon_k) \, \delta_{kk'} \, , \label{limith} \\
\lim_{\underset{\langle A  \rangle  = \text{A}}{\text{T}\rightarrow 0 \, \& \, \Delta \rightarrow 0}} \kappa_{kk'} & = \lim_{\underset{\langle A  \rangle  = \text{A}}{\text{T}\rightarrow 0 \, \& \, \Delta \rightarrow 0}} u_k v_k \, \delta_{\bar{k}k'} \nonumber \\
&= 0  \, , 
\end{align}
\end{subequations}
where $\Theta(x)$ denotes the Heaviside function. As a result, hole (particle) states are occupied with probability $1$ ($0$). 

Consequently, the HFB state converges trivially to the HF closed-shell Slater determinant defined in Eq.~\eqref{HFslatervacuum}
\begin{align}
\lim_{\underset{\langle A  \rangle  = \text{A}}{\text{T}\rightarrow 0 \, \& \, \Delta \rightarrow 0}}   | \Phi \rangle & =  | \bar{\Phi} \rangle =  \prod_{h=1}^{A/2} a^{\dagger}_h a^{\dagger}_{\bar{h}}| 0 \rangle \, , \label{HFBstatelimit0}
\end{align}
which is an eigenstate of $A$ with zero particle-number variance
\begin{align}
\lim_{\underset{\langle A  \rangle  = \text{A}}{\text{T}\rightarrow 0 \, \& \, \Delta \rightarrow 0}} \Delta \text{A}_{D} &= 0 \, . \label{HFBvariancelimit1}
\end{align}
The energy takes the standard mean-field form associated with a Slater determinant 
\begin{align}
\lim_{\underset{\langle A  \rangle  = \text{A}}{\text{T}\rightarrow 0 \, \& \, \Delta \rightarrow 0}} \text{E}_{D} &= \sum_{h=1}^{\text{A}} t_{hh}  + \frac{1}{2} \sum_{hh'=1}^{\text{A}} \overline{v}_{hh'hh'}  \, . \label{HFBenergylimit1}
\end{align}
In summary, no surprise occurs for closed-shell systems in the $\text{T}\rightarrow 0 \, \& \, \Delta \rightarrow 0$ limits; i.e. their description is given by the HF closed-shell Slater determinant.

\subsubsection{Open-shell system} 

As discussed in  Ref.~\cite{Duguet:2020hdm}, a non-trivial solution is obtained for open-shell systems. While hole and particle states still behave according to Eq.~\eqref{1bodydensmatlimits} such that $\text{A}-a_v$ particles eventually occupy hole states, $a_v$ particles need to be placed into the $d_v$ ($p_v=d_v/2$ pairs of) degenerate valence states characterized by equal occupations\footnote{As pointed out in Ref.~\cite{Duguet:2020hdm}, the half-filled shell ($o_v = 1/2$) must be treated with extra care. Thus, we consider that $o_v \neq 1/2$ for simplicity in the present section.}  $0<o_v<1$. Achieving this requires that $|\epsilon_{v} - \lambda|$ and $\Delta_{v}$ go to 0 in a strictly proportional fashion, i.e.
\begin{equation}
\lim_{\underset{\langle A  \rangle  = \text{A}}{\text{T}\rightarrow 0 \, \& \, \Delta \rightarrow 0}} \left|\frac{\Delta_{v}}{\epsilon_{v} - \lambda}\right| = \frac{2\sqrt{o_v(1-o_v)}}{|1- 2 o_v|}  \, , \label{limit7}
\end{equation}
as was  numerically illustrated in Ref.~\cite{Duguet:2020hdm}. In this context, the one-body density matrices given in Eq.~\eqref{1bodydensmatlimits} within the particle and hole subspaces are complemented within the valence shell by
\begin{subequations}
\begin{align}
\lim_{\underset{\langle A  \rangle  = \text{A}}{\text{T}\rightarrow 0 \, \& \, \Delta \rightarrow 0}} \rho_{v_{k}v_{k'}} & = o_v \, \delta_{kk'} \, , \\
\lim_{\underset{\langle A  \rangle  = \text{A}}{\text{T}\rightarrow 0 \, \& \, \Delta \rightarrow 0}} \kappa_{v_{k}v_{k'}} & =  \sqrt{o_v(1-o_v)}  \, \delta_{\bar{k}k'}\, , 
\end{align}
\end{subequations}
whereas the reference HFB state becomes
\begin{align}
\lim_{\underset{\langle A  \rangle  = \text{A}}{\text{T}\rightarrow 0 \, \& \, \Delta \rightarrow 0}} | \Phi \rangle &=  \prod_{k=1}^{p_v} (\sqrt{1-o_v} + \sqrt{o_v} \, a^{\dagger}_{v_{k}} a^{\dagger}_{v_{\bar{k}}})| \tilde{\Phi} \rangle \, , \label{HFBstatelimit2}
\end{align}
where $| \tilde{\Phi} \rangle$ was introduced in Eq.~\eqref{coreSD}. Thus, the even-number-parity HFB state carrying A particles on average becomes a linear combination of $2^{p_v}$ Slater determinants among which $\binom{b}{p_v}$ of them carry $B(b)=\text{A}-a_v+2b$ particles, with the integer $b$ running from $0$ to $p_v$. The even number of particles carried by the Slater determinants thus ranges from $\text{A}-a_v$ to $\text{A}+ (d_v-a_v)$.

While the entropy vanishes in the present case, the particle-number variance does not. It evaluates to
\begin{align}
\lim_{\underset{\langle A  \rangle  = \text{A}}{\text{T}\rightarrow 0 \, \& \, \Delta \rightarrow 0}} \Delta \text{A}_{D} &= \  2 a_v (1-o_v)\,  \label{varGEN}
\end{align}
and constitutes a lower bound within the manifold of even-number-parity HFB states carrying A particles on average~\cite{Duguet:2020hdm}. Furthermore, the binding energy contains a non-zero pairing contribution of the form
\begin{align}
\lim_{\underset{\langle A  \rangle  = \text{A}}{\text{T}\rightarrow 0 \, \& \, \Delta \rightarrow 0}} E^{\text{B}}_{D}  & =  o_v(1-o_v)\sum_{kl=1}^{p_v}
\overline{v}_{v_{k}v_{\bar{k}}v_{l}v_{\bar{l}}}  \, . \label{HFBenergylimit3}
\end{align}

The above analysis demonstrates that HFB theory does {\it not} reduce to the HF formalism for open-shell systems when the pairing field is driven to zero in the HFB Hamiltonian matrix~\cite{Duguet:2020hdm}. 

\subsection{$\Delta\rightarrow 0 \, \& \, \text{T}\rightarrow 0$}
\label{case_2b}

The main objective of the present work is to investigate case 2.b, i.e. the case in which both limits are taken in the opposite order. Based on the result of Sec.~\ref{case_1b} above, this case boils down to taking the zero-temperature limit of the FTHF formalism. The main outcomes of the analysis given below are that the resulting description of open-shell systems is
\begin{enumerate}
\item non trivial, i.e. does {\it not} reduce to the straight HF formalism,
\item different from case 2.a, i.e. the zero-pairing and zero-temperature limits do {\it not} commute.
\end{enumerate}
In order to make the situation transparent, closed-shell and open-shell systems must again be distinguished.

\subsubsection{Closed-shell system} 

In a closed-shell system\footnote{We remind the reader that we do not restrict the notion of ``closed-shell'' to the spherically symmetric case. The notion refers only to the occupations of the nuclear levels in the combined zero-temperature and zero-pairing limits, such that a system can be ``deformed closed-shell''. } where nuclear shells strictly separate into A hole ($\epsilon_h - \lambda<0$) states and the remaining particle ($\epsilon_p - \lambda>0$) states, the statistical occupations (Eq.~\eqref{stat_weight_zeropairing}) become
\begin{align}
\lim_{\underset{\text{A}_{D}  = \text{A}}{\Delta\rightarrow 0 \, \& \, \text{T}\rightarrow 0}}  \zeta^{-1}_{h} = \lim_{\underset{\text{A}_{D}  = \text{A}}{\Delta\rightarrow 0 \, \& \, \text{T}\rightarrow 0}}  \zeta_{p} &= 0   \, . \label{limitstatoccup}
\end{align}
Consequently, the zero-temperature limit of the FTHF density operator (Eqs.~\eqref{explicitD_zeropairing}-\eqref{explicitZ_zeropairing}) is nothing but a {\it pure state}
\begin{align}
\lim_{\underset{\text{A}_{D} = \text{A}}{\Delta\rightarrow 0 \, \& \, \text{T}\rightarrow 0}} D &= | \bar{\Phi} \rangle \langle \bar{\Phi} | \, ,
\end{align}
where $| \bar{\Phi} \rangle$ denotes the closed-shell HF Slater determinant carrying A nucleons (Eq.~\eqref{HFslatervacuum}). Thus, the entropy becomes zero and the normal one-body density matrix (Eq.~\eqref{limitdensmatrices_zero_pairing1}) reduces to
\begin{align}
\lim_{\underset{\text{A}_{D}  = \text{A}}{\Delta\rightarrow 0 \, \& \, \text{T}\rightarrow 0}} \rho_{kk'} &=  \Theta(\lambda - \epsilon_k) \delta_{kk'}  \, , \label{limitdensmatrix_zero_temp}
\end{align}
such that hole (particle) states are occupied with probability $1$ ($0$). The particle-number variance is also zero whereas the energy is given by Eq.~\eqref{HFBenergylimit1}. 

In summary, no surprise occurs for closed-shell systems. The description is given by the HF closed-shell Slater determinant and is identical to the $\text{T}\rightarrow 0 \, \& \, \Delta \rightarrow 0$ case, i.e. the description is independent of the order in which both limits are performed. 

\subsubsection{Open-shell system} 

Let us now study the zero-temperature limit of the FTHF formalism for an open-shell system. The occupation of hole and particle states behaves as for closed-shell systems, i.e. 
\begin{align}
\lim_{\underset{\text{A}_{D}  = \text{A}}{\Delta\rightarrow 0 \, \& \, \text{T}\rightarrow 0}} \rho_{kk'} &=  \Theta(\lambda - \epsilon_k) \, \delta_{kk'}  \, , \label{limitdensmatrix_zero_temppair}
\end{align}
for $k=h$ or $p$, such that $\text{A}-a_v$ particles eventually occupy hole states with probability $1$. The valence shell needs to fit the remaining $a_v$ particles in its $d_v$ degenerate states, which requires that 
\begin{align}
\lim_{\underset{\text{A}_{D}  = \text{A}}{\Delta\rightarrow 0 \, \& \, \text{T}\rightarrow 0}} \rho_{v_kv_k'} &=  o_v  \, \delta_{kk'} \, , \label{limitfv_zero_temp}
\end{align}
for $(k,k')\in [1,d_v]^2$. Given Eq.~\eqref{limitdensmatrices_zero_pairing1}, satisfying Eq.~\eqref{limitfv_zero_temp} necessarily implies that
\begin{align}
\lim_{\underset{\text{A}_{D}  = \text{A}}{\Delta\rightarrow 0 \, \& \, \text{T}\rightarrow 0}} \frac{\epsilon_{v} - \lambda}{\text{T}} &=    \ln{\left(\frac{1-o_v}{o_v}\right)} \equiv \gamma_v \, . \label{limit7B}
\end{align}
For $o_v \neq  1/2$, this means that $\epsilon_{v} - \lambda$ and $T$ must go to 0 in a strictly proportional fashion. For $o_v =  1/2$, $\epsilon_{v} - \lambda$ goes to 0 faster than $T$. 

In this situation, the zero-temperature limit of the FTHF density operator defined through Eqs.~\eqref{explicitD_zeropairing}-\eqref{explicitZ_zeropairing} takes the non-trivial form
\begin{align}
\lim_{\underset{\text{A}_{D}  = \text{A}}{\Delta\rightarrow 0 \, \& \, \text{T}\rightarrow 0}}    D &= \frac{1}{Z_{3}} e^{-\gamma_v \sum_{k \in [1,d_v]}  a^{\dagger}_{v_k} a_{v_k}} \, ,
    \label{FTHFB_densityoperator_zeropairingT}
\end{align}
where
\begin{align}
    Z_{3} &\equiv  \Tr  \left[e^{-\gamma_v \sum_{k \in [1,d_v]}  a^{\dagger}_{v_k} a_{v_k}}\right] \, . \label{Z3}
\end{align}
Contrarily to the closed-shell case, the FTHF density operator describing an open-shell system is {\it not} associated with a pure state in the zero-temperature limit. Given the specific form of the one-body density matrix (Eqs.~\eqref{limitdensmatrix_zero_temppair}-\eqref{limitfv_zero_temp}), the observables of interest are easily computed from Eq.~\eqref{eq:calc_limdelta} in the present limit. In particular, the particle-number variance is non-zero and equal to 
\begin{align}
\lim_{\underset{\text{A}_{D}  = \text{A}}{\Delta\rightarrow 0 \, \& \, \text{T}\rightarrow 0}}  \Delta \text{A}_{D}  &= \lim_{\underset{\text{A}_{D}  = \text{A}}{\Delta\rightarrow 0 \, \& \, \text{T}\rightarrow 0}} \tr \left[ \rho (1 - \rho) \right]  \nonumber \\
               &= \sum_{k=1}^{d_v} \rho_{v_kv_k} (1-\rho_{v_kv_k}) \nonumber \\
               &= a_v(1-o_v)  \, , \label{eq:calcDeltaNlimdeltaT} 
\end{align}
whereas the entropy
\begin{align}
\lim_{\underset{\text{A}_{D}  = \text{A}}{\Delta\rightarrow 0 \, \& \, \text{T}\rightarrow 0}} \text{S}_{D} &=  - d_{\nu} \left[ o_{\nu} \ln o_{\nu} + (1-o_{\nu}) \ln \left( 1-o_{\nu} \right) \right]
\label{entropylimits}
\end{align}
is also non-zero in spite of the zero-temperature limit. One observes that the particle-number variance is {\it half} of the one obtained when performing the limits in the alternative order (Eq.~\eqref{varGEN}). In the present case the residual particle-number fluctuation has a thermal origin as reflected by the entropy. In the $\text{T}\rightarrow 0 \, \& \, \Delta \rightarrow 0$ case, the number fluctuation is instead generated by lingering pairing correlations as reflected by the non-zero anomalous density matrix.

Let us further explore the structure of the density operator by expanding it explicitly on the basis of Fock space built out of the closed-shell Slater determinant $| \tilde{\Phi} \rangle$ carrying  $\text{A}-a_v$  (Eq.~\eqref{coreSD}). Given Eqs.~\eqref{limitstatoccup} and~\eqref{FTHFB_densityoperator_zeropairingT}-\eqref{Z3}, Eq.~\eqref{commut_D_zeropairing} provides in the present case
\begin{subequations}
\label{commut_D_zeropairingT}
\begin{align}
\lim_{\underset{\text{A}_{D}  = \text{A}}{\Delta\rightarrow 0 \, \& \, \text{T}\rightarrow 0}}    D|  \tilde{\Phi} \rangle &= \frac{1}{Z_{3}} |  \tilde{\Phi} \rangle \, , \label{D_vac_zeropairingT} \\
\lim_{\underset{\text{A}_{D}  = \text{A}}{\Delta\rightarrow 0 \, \& \, \text{T}\rightarrow 0}}  D a^{\dagger}_{p} D^{-1} &= 0 \, , \label{commut_D_crea_p_zeropairingT} \\
\lim_{\underset{\text{A}_{D}  = \text{A}}{\Delta\rightarrow 0 \, \& \, \text{T}\rightarrow 0}}   D a_{h} D^{-1} &= 0  \, , \label{commut_D_anni_h_zeropairingT} \\
\lim_{\underset{\text{A}_{D}  = \text{A}}{\Delta\rightarrow 0 \, \& \, \text{T}\rightarrow 0}}  D a^{\dagger}_{v_k} D^{-1} &= e^{-\gamma_v} a^{\dagger}_{v_k} \, , \label{commut_D_crea_v_zeropairingT} \\
\lim_{\underset{\text{A}_{D}  = \text{A}}{\Delta\rightarrow 0 \, \& \, \text{T}\rightarrow 0}}   D a_{v_k} D^{-1} &=  e^{+\gamma_v}  a_{v_k}  \, , \label{commut_D_anni_v_zeropairingT} 
\end{align}
\end{subequations}
As a result, Eqs.~\eqref{explicitD_zeropairing}-\eqref{explicitZ_zeropairing} are transformed into
\begin{align}
    \lim_{\underset{\text{A}_{D}  = \text{A}}{\Delta\rightarrow 0 \, \& \, \text{T}\rightarrow 0}} D  &=\frac{1}{Z_3}\Big(| \tilde{\Phi} \rangle \langle \tilde{\Phi} | \nonumber \\
    &\hspace{1cm} + s_v \sum_{k \in [1,d_v]}| \tilde{\Phi}^{v_k} \rangle \langle \tilde{\Phi}^{v_k} | \nonumber \\
    &\hspace{1cm} + \frac{(s_v)^2}{2!} \sum_{(k,k') \in [1,d_v]^2}  | \tilde{\Phi}^{v_k v_{k'}} \rangle \langle \tilde{\Phi}^{v_k v_{k'}} |  \nonumber \\
    &\hspace{1cm} \vdots \nonumber \\
    &\hspace{1cm} + (s_v)^{d_v}  | \tilde{\Phi}^{v_1 \ldots v_{d_v}} \rangle \langle \tilde{\Phi}^{v_1 \ldots v_{d_v}} |  \Big) \, , \label{explicitDlimits}
\end{align}
with 
\begin{align}
    Z_3 &= \left(1+ s_{v}\right)^{d_v}  \, , \label{explicitZlimits}
\end{align}
where the definition
\begin{align}
s_v \equiv e^{-\gamma_v} = \frac{1-o_v}{o_v}  \label{limitxiv_zero_temp2}
\end{align}
has been introduced. Equation~\eqref{explicitDlimits} involves the $\binom{b}{d_v}$ Slater determinants built from $| \tilde{\Phi} \rangle$ by creating $b$  ($b=0,1,\ldots,d_v$) particles in the valence shell
\begin{align}
| \tilde{\Phi}^{v_k \ldots v_{k'}} \rangle & \equiv a^{\dagger}_{v_{k}} \ldots a^{\dagger}_{v_{k'}} | \tilde{\Phi} \rangle \, , \label{excitedcoreSD}
\end{align}
with $(k,\ldots k') \in [1,d_v]^{b}$. The FTHF density operator is thus a statistical mixture of $\sum_{b=0}^{d_v} \binom{b}{d_v} = 2^{d_v}$ Slater determinants whose {\it even or odd} number of particles vary from $\text{A}-a_v$ to $\text{A}+ (d_v-a_v)$. 

Illustratively, the average particle number and the particle-number variance (already computed in Eq.~\eqref{eq:calcDeltaNlimdeltaT} through traces in the one-body Hilbert space) can be recovered through traces in Fock space  on the basis of Eqs.~\eqref{explicitDlimits}-\eqref{explicitZlimits}, i.e.\footnote{Identities~\eqref{binomial1} and~\eqref{binomial2} provided in App.~\ref{formulae} are employed to derive Eq.~\eqref{averGEN2} while the additional identity~\eqref{binomial3} is necessary to derive Eq.~\eqref{varGEN2}. Similar analytical results can be derived for higher moments of $A$ by considering higher derivatives of Newton's binomial formula.}
\begin{subequations}
\begin{align}
\lim_{\underset{\text{A}_{D}  = \text{A}}{\Delta\rightarrow 0 \, \& \, \text{T}\rightarrow 0}}  \text{A}_{D}  &= \lim_{\underset{\text{A}_{D}  = \text{A}}{\Delta\rightarrow 0 \, \& \, \text{T}\rightarrow 0}} \Tr \left[ D A     \right]  \nonumber \\
               &= \frac{1}{(1+s_v)^{d_v}} \sum_{b=0}^{d_v} \binom{b}{d_v} s^{b}_v (\text{A}-a_v+b)\nonumber \\
&= \text{A} \, , \label{averGEN2} \\
\lim_{\underset{\text{A}_{D}  = \text{A}}{\Delta\rightarrow 0 \, \& \, \text{T}\rightarrow 0}}  \Delta \text{A}_{D}  &= \lim_{\underset{\text{A}_{D}  = \text{A}}{\Delta\rightarrow 0 \, \& \, \text{T}\rightarrow 0}} \Tr \left[ D A^2 \right] -  \Tr \left[D A\right]^2\nonumber \\
&= \frac{1}{(1+s_v)^{d_v}} \sum_{b=0}^{d_v} \binom{b}{d_v} s^{b}_v (\text{A}-a_v+b)^2 \nonumber \\
&\phantom{=} - \text{A}^2 \nonumber \\
&=   a_v (1-o_v)\, , \label{varGEN2}
\end{align}
\end{subequations}
where the latter result indeed agrees with Eq.~\eqref{eq:calcDeltaNlimdeltaT}.

\subsection{Discussion}

Let us further comment on a few key points.
\begin{itemize}
\item  In Ref.~\cite{Duguet:2020hdm}, HFB theory was shown {\it not} to reduce {\it in all cases} to HF theory when the pairing field is driven to zero. Similarly, the above analysis demonstrates that FTHF theory does {\it not} reduce to the straight HF formalism  {\it in all cases} when the temperature is driven to zero. 
 \item Starting from FTHFB theory, the description of open-shell systems in the combined zero-temperature and zero-pairing limits is shown to depend on the order with which both limits are taken, i.e. the description is either given by a pure state made out of a linear combination of a finite number of Slater determinants with even particle numbers or by a statistical mixture of a finite number of Slater determinants with both even and odd particle numbers. 
 \item This difference in the obtained many-body description leads to unexpected expectation values of operators. Above, the particle-number variance was shown to be non-zero and to differ by a factor of 2 in both cases. Such a feature is not limited to the particle-number variance but extends to any operator involving a product of more than one creation or annihilation operator. 
\item It has been standard to perform HF calculations of open-shell nuclei within the so-called equal filling approximation (EFA), equally distributing the valence nucleons among the levels in a valence shell to guarantee spherical symmetry. For a long time though, this procedure was typically applied without any formal justification. In Ref.~\cite{PerezMartin:2008yv}, a first justification of the EFA procedure was delivered on the basis of a specifically-tuned ensemble-HFB theory. In Ref.~\cite{Duguet:2020hdm}, a second justification was provided in terms of a pure state obtained via the zero-pairing-limit of straight HFB theory (i.e. the $\text{T}\rightarrow 0 \, \& \, \Delta\rightarrow 0$ limit of the present work). Eventually, a third justification of the EFA is presently given in terms of the statistical
mixture obtained through the $\Delta\rightarrow 0 \, \& \, \text{T}\rightarrow 0$ limit of FTHFB theory\footnote{The ensemble HFB theory designed in Ref.~\cite{PerezMartin:2008yv} is more general as it justifies the EFA in presence of pairing correlations. It is particularly suited to include the blocking effect associated with the unpaired odd particle when dealing with odd systems. In the zero-pairing limit, the statistical mixture does coincide with the one obtained presently.}.

\end{itemize}

\section{Applications}
\label{sec:numerical}

In this section, results obtained from constrained FTHFB calculations are presented to illustrate the findings of the previous sections.

\subsection{Numerical set up}
\label{sec:num}

Following Ref.~\cite{Duguet:2020hdm}, the three semi-magic oxygen isotopes $^{18,22,26}$O are employed as test cases of FTHFB calculations in which time-reversal and rotational invariances are imposed. While $^{18}$O ($o_{d_{5/2}} = 1/3$) and $^{26}$O ($o_{d_{3/2}} = 1/2$) are representative of spherical (neutron) open-shell nuclei, $^{22}$O  ($o_{d_{5/2}} = 1$) is the token closed-shell system. We further consider $^{19}$O ($o_{d_{5/2}} = 1/2$) as an example of an odd nucleus, albeit calculated without breaking time-reversal symmetry. Finite-temperature HFB calculations are performed within the sd valence space on the basis of the standard USD interaction~\cite{Wildenthal1984,Brown1988}. Oxygen isotopes are described as having an inert core of $^{16}$O, implying that the protons play no role and that $^{18}$O, $^{19}$O, $^{22}$O and $^{26}$O possess $2,3,6$ and $10$ active valence neutrons, respectively. The working equations are solved on the basis of the HF-SHELL code~\cite{Ryssens2020}. In the present context, virtually all sd-shell nuclei exhibit a deformed mean-field minimum and several of them are triaxial~\cite{Stetcu2002}. While HF-SHELL is sufficiently general to study triaxial nuclear shapes, the calculations here have been restricted to spherical configurations except if specified otherwise\footnote{All spherical solutions obtained here, with the exception of the spherical closed-shell $^{22}$O, are saddle points with respect to quadrupole deformation. If deformed solutions were to be authorized, $^{18}$O, $^{19}$O and $^{26}$O would actually qualify as {\it deformed} closed-shell systems in the zero-pairing and/or zero-temperature limits. The calculations are thus restricted to spherical symmetry to illustrate the behavior of the formalism in these limits whenever the system is (constrained to be) of open-shell character.}. This is in practice achieved by initializing the iterative process with a perfectly spherical state.

\subsection{Characterization of the combined limits}

For the $\text{T}\rightarrow 0 \, \& \, \Delta\rightarrow 0$ limit to be analytically meaningful in open-shell systems, canonical matrix elements of the pairing field were predicted to be driven to zero in a specific manner (Eq.~\eqref{limit7}) when the constraining parameter $\delta$ goes itself to zero. This key feature was confirmed numerically in Ref.~\cite{Duguet:2020hdm} and is thus not repeated here.

\begin{figure}
    \centering
    \includegraphics[width=.5\textwidth]{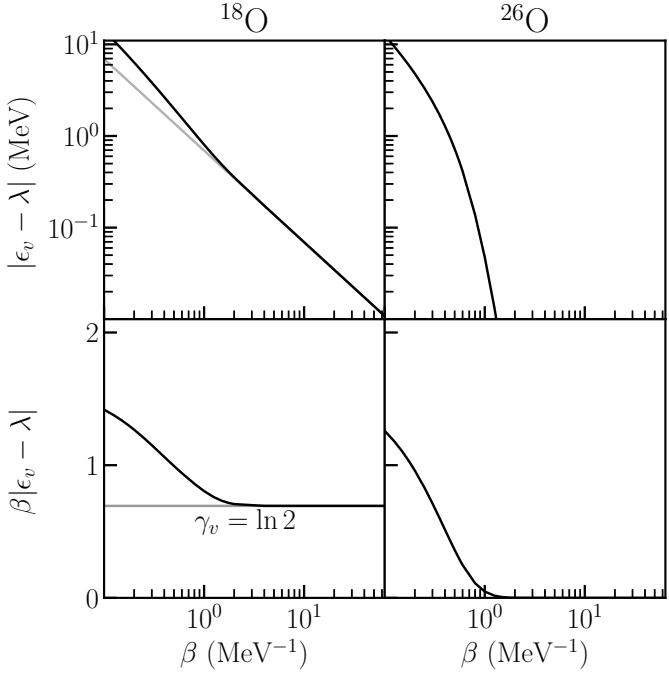}
    \caption{Results of FTHFB calculations for $^{18}$O (left column) and $^{26}$O (right column) in the zero-pairing limit as a function of the inverse temperature. Top row: $|\epsilon_{v} - \lambda|$. Bottom row: $\beta |\epsilon_{v} - \lambda|$. Faint gray lines indicate the values predicted analytically (Eq.~\eqref{limit7B}) in the zero-temperature limit.}
    \label{fig:limitcharacterization}
\end{figure}

For the $\Delta\rightarrow 0 \, \& \, \text{T}\rightarrow 0$ limit to be analytically meaningful in open-shell systems, the FTHF quasi-particle energy $|\epsilon_{v} - \lambda|$ associated with the valence shell has been predicted to be driven to zero in a specific way (Eq.~\eqref{limit7B}) when the temperature goes itself to zero. The top panels of Fig.~\ref{fig:limitcharacterization} display $|\epsilon_{v} - \lambda|$ against the inverse temperature $\beta$ for $^{18}$O and $^{26}$O. In agreement with  Eq.~\eqref{limit7B}, $|\epsilon_{v} - \lambda|$ goes to zero strictly proportionally to (faster than) $\text{T}$ in $^{18}$O ($^{26}$O) whose valence shell occupation $o_v$ is different from (equal to) $1/2$. Going one step further, the product $\beta|\epsilon_{v} - \lambda|$, which is analytically predicted to converge to a characteristic value in the zero-temperature limit according to Eq.~\eqref{limit7B}, is displayed in the bottom panels of Fig.~\ref{fig:limitcharacterization}. The predicted limit is accurately obtained numerically for both systems.

\subsection{Particle-number variance}

\begin{figure}
    \centering
    \includegraphics[width=.5\textwidth]{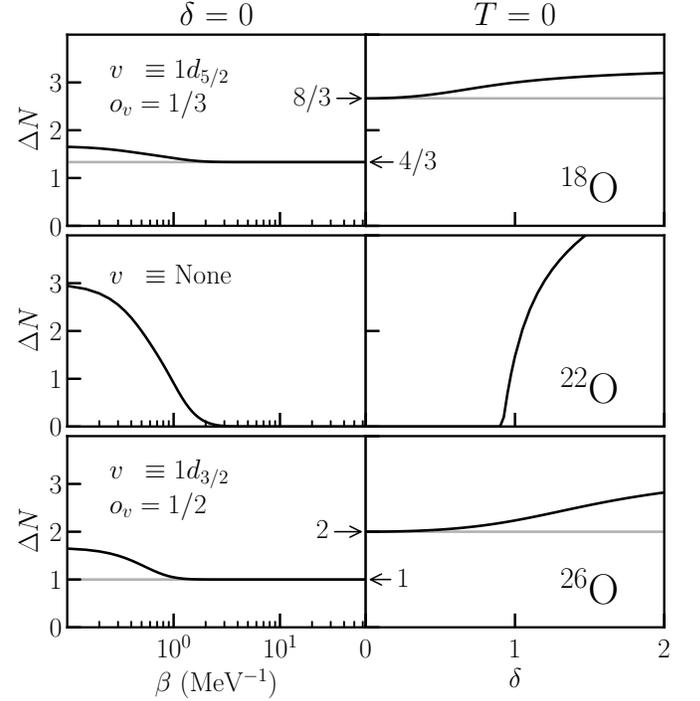}
    \caption{Finite-temperature HFB neutron-number variance in $^{18}$O (top row), $^{22}$O (middle row) and $^{26}$O (bottom row). Left column: results as a function the effective pairing strength $\delta$ at zero temperature. Right column: results as a function the inverse temperature $\beta$ at zero effective pairing strength. Faint gray lines indicate the limiting values as predicted by Eq.~\eqref{varGEN2} (left) and Eq.~\eqref{varGEN} (right). }
    \label{fig:dispersion}
\end{figure}

With the aim to further characterize the FTHFB density operator in the combined zero-temperature and zero-pairing limits, the neutron-number variance is displayed in the right (left) column of Fig.~\ref{fig:dispersion} for $^{18,22,26}$O as a function of the effective pairing strength $\delta$ (inverse temperature $\beta$) at $\text{T}=0$ ($\delta =0$). 

One first observes that the results displayed in the right column of Fig.~\ref{fig:dispersion} are consistent with those reported in Ref.~\cite{Duguet:2020hdm}. In particular, the limit values obtained for $\delta \rightarrow 0$ agree in all cases with the prediction of Eq.~\eqref{varGEN}\footnote{The fact that the neutron-number variances obtained at large values of $\delta$ are significantly lower than those obtained in Ref.~\cite{Duguet:2020hdm} is simply due to the smaller single-particle model space employed here.}. As for the left column, one also observes that the dispersion monotonically decreases with decreasing temperature and reaches in all three nuclei the value predicted through Eq.~\eqref{varGEN2}.

For closed-shell $^{22}$O, the neutron-number variance goes to zero in both cases such that the same value is obtained independently of the order with which the two limits are performed. While it does reach zero smoothly in the left column as the temperature decreases, the neutron-number variance suddenly drops to zero for a non-zero value of the effective pairing strength $\delta$ in the right column, which reflects the well-known pairing collapse occurring in straight HFB theory. 

In open-shell nuclei $^{18}$O and $^{26}$O, the neutron-number variance goes to non-zero values in both columns. Furthermore, the limit values are consistent with the theoretical predictions (Eqs.~\eqref{varGEN} and~\eqref{varGEN2}). In particular, one observes that the zero-pairing and zero-temperature limits do not commute such that\footnote{As a consequence of the finite dimension of the model space, the maximum particle-number variance achievable at high temperature is also half of the maximum value achievable at large $\delta$.}
\begin{align}
\lim_{\underset{\text{A}_{D}  = \text{A}}{\text{T} \rightarrow 0 \, \& \, \Delta\rightarrow 0}}  \Delta \text{A}_{D} &= 2 \lim_{\underset{\text{A}_{D}  = \text{A}}{\Delta\rightarrow 0 \, \& \, \text{T}\rightarrow 0}}  \Delta \text{A}_{D} \, , 
\end{align}
in these two nuclei. Clearly, the neutron-number variance obtained in the combined limits acts in all cases as a (possibly non-zero) lower bound  within the manifold of appropriate FTHFB density operators.  

\begin{figure}
    \centering
    \includegraphics[width=.5\textwidth]{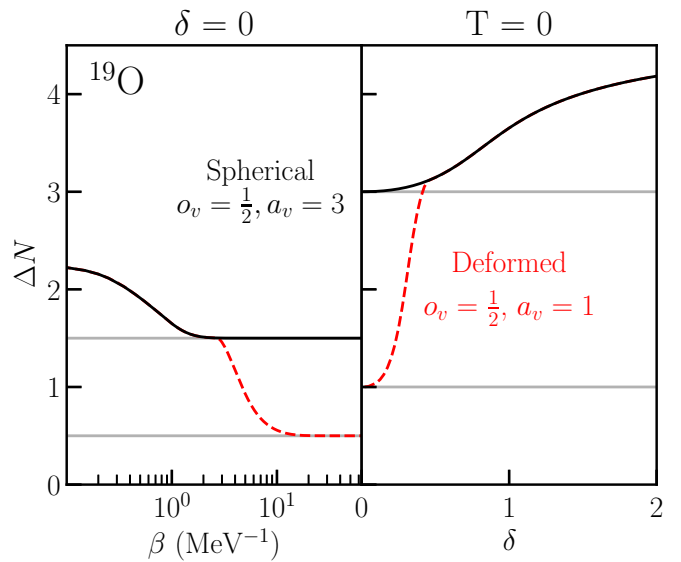}
    \caption{Same as Fig.~\ref{fig:dispersion} for $^{19}$O calculated as a false vacuum~\cite{Duguet:2001gr}. The black full line follows the spherical configuration ($o_v = 1/2, a_v = 3$). Around $\beta \sim 3 $ MeV (left) or $\delta \sim 0.4$ (right), the spherical solution becomes a saddle point. The neutron-number variance of the oblate solution ($o_v = 1/2, a_v = 1$) is indicated by the red dashed line.  Faint gray lines indicate the limiting values of the 
    neutron-number variances predicted analytically in each case.}
    \label{fig:dispersion2}
\end{figure}

Figure~\ref{fig:dispersion2} displays the neutron-number variance for yet another case of interest, i.e. the odd isotope $^{19}$O calculated as a fully-paired vacuum~\cite{Duguet:2001gr}. This hypothesis consists of treating the odd system while imposing (at least) one symmetry, i.e. time reversal invariance\footnote{In symmetry-unrestricted calculations, the description of odd systems requires time-reversal symmetry to be broken. As a result, canonical nuclear shells do not display any degeneracy such that the odd nucleus eventually converges to a closed-shell configuration in the joint zero-pairing and zero-temperature limits.}. Furthermore, the calculation is performed twice to illustrate the effect of relaxing one (but not all) of the symmetry restrictions, i.e. a first time constraining the solution to spherical symmetry and a second time authorizing it to break rotational invariance. Because the average particle number is constrained to an odd value and the one-body density matrices manifest time-reversal symmetry, the system is constrained to manifest an open-shell character in the joint limits. Even when the $2j+1$-fold degeneracy of single-particle shells associated with spherical symmetry is lifted through deformation, the remaining two-fold Kramer's degeneracy imposes that the naive occupation of the valence shell is necessarily $o_v=1/2$. 

The solution of $^{19}$O constrained to spherical symmetry is characterized by $o_v = 1/2$ and $a_v = 3$. The neutron-number variance thus reaches, in agreement with  Eqs.~\eqref{varGEN} and~\eqref{varGEN2}, $3$ and $3/2$ depending on the order of the limits. When authorizing the system to deform, the system has energetic advantage to do so such that the naive filling of the (two-fold degenerate) valence shell is characterized by $o_v = 1/2$ and $a_v = 1$. As a result, the particle number variance of the oblate solution converges, in agreement with  Eqs.~\eqref{varGEN} and~\eqref{varGEN2}, to either $1$ or $1/2$. The solutions obtained in the $\Delta\rightarrow 0 \, \& \, \text{T}\rightarrow 0$ limits correspond to the spherical or deformed HF-EFA approximation. Contrarily, and as already pointed out in Ref.~\cite{Duguet:2020hdm}, the solutions obtained in the  $\text{T}\rightarrow 0 \, \& \, \Delta\rightarrow 0$ limits do not correspond to the EFA but still constitute solutions found for an open-shell configuration in the zero-pairing limit of straight HFB.

\subsection{Spectroscopic quantities}

\begin{figure}
    \centering
    \includegraphics[width=.47\textwidth]{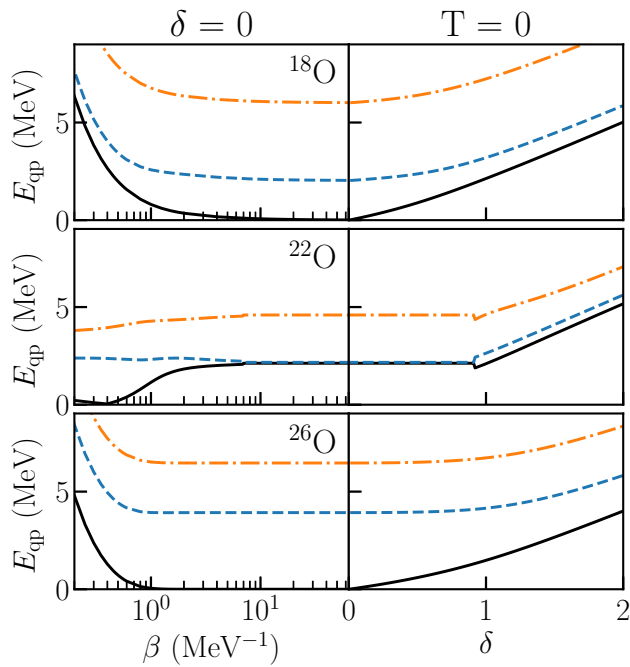}
    \caption{(Color online) Same as Fig.~\ref{fig:dispersion} for the three lowest quasi-neutron energies. The (online) colors
stipulate the first (black full line), second (blue dashed line) and third (orange dash-dotted line) lowest quasi-neutron energies. The discontinuity appearing in the right column for $^{22}$O signals the pairing collapse. }
    \label{fig:neutronqp}
\end{figure}

The FTHFB density operator reached in the combined zero-pairing and zero-temperature limits is further scrutinized in Fig.~\ref{fig:neutronqp} where the three lowest quasi-neutron energies $E_k$ (Eq.~\eqref{eq:Hhfbdiag}) are displayed in $^{18,22,26}$O as a function the effective pairing strength $\delta$ (inverse temperature $\beta$) at $\text{T}=0$ ($\delta =0$). 

One observes that quasi-neutron energies, while behaving differently in the left and right columns as a function of the inverse temperature and effective pairing strength, respectively, reach the same values independently on the order with which the limits are performed. While this result is trivial in closed-shell nuclei for which the two limits commute, it is not in open-shell nuclei. In fact, the density operators obtained in the two orderings display identical normal one-body density matrices and only differ through the anomalous density matrix (within the valence shell) entering the Bogoliubov field. Since the pairing field is anyway driven to zero through the zero-pairing limit, the HFB matrix is identical in both cases, and so are its eigenvalues, i.e. independently of the order with which the limits are taken, one has that $E_k \rightarrow |\epsilon_k - \lambda|$. We emphasize that, while the HFB matrix is identical in both cases, the many-body description constructed with the help of its eigenvectors is not.

Eventually, while the lowest quasi-neutron energy remains non-zero in $^{22}$O, it goes to zero in open-shell nuclei as testimony of the fact that $|\epsilon_v - \lambda| \rightarrow 0$ in the combined limits. The consequences of this degeneracy of open-shell ground states with respect to elementary excitations in beyond mean-field methods such as Bogoliubov many-body perturbation theory~\cite{Duguet:2015yle,Tichai:2018vjc,Arthuis:2018yoo,Tichai:2020dna} was illustrated in Ref.~\cite{Duguet:2020hdm}.

\subsection{Grand potential}

\begin{figure*}
    \centering
    \includegraphics[width=.88\textwidth]{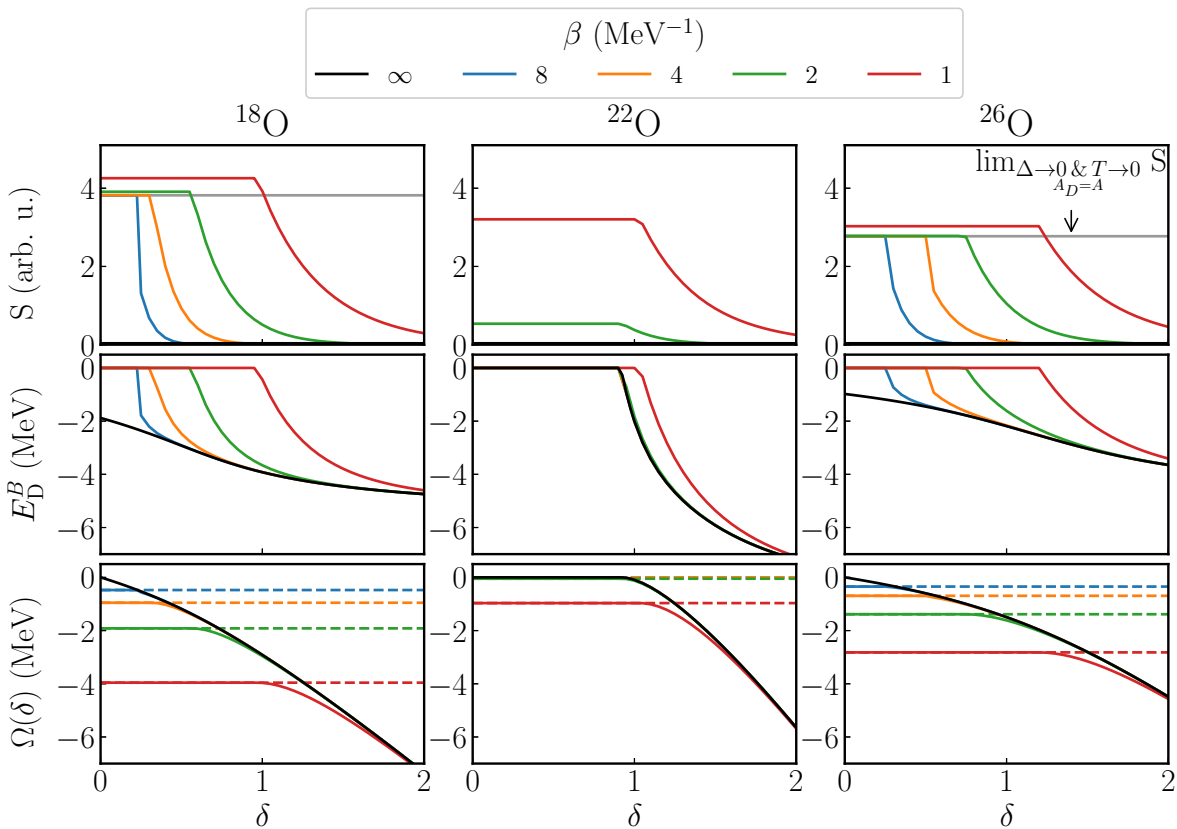}
    \caption{(Color online) Finite-temperature HFB entropy (top row), pairing energy (middle row) and Routhian $\Omega(\delta)_{D}$ (bottom) as a function of the effective pairing strength $\delta$ for different values of the inverse temperature in $^{18}$O (left column), $^{22}$O (center column) and $^{26}$O (right column). The Routhian is normalized to the ground-state result at $\delta = 0$ for each isotope. The black lines ($\beta = \infty$) denote the result of ground-state HFB calculations. Gray lines in the top row indicate the limiting values of the entropy in the combined $\Delta\rightarrow 0 \, \& \, \text{T}\rightarrow 0$ limits (Eq.~\eqref{entropylimits}). The dashed lines in the bottom row indicate the value of the Routhian for the FTHF calculation at corresponding temperature.}
    \label{fig:entropy}
\end{figure*}

Figure~\ref{fig:entropy} illustrates the competition between the pairing energy and the entropy in the constrained grand potential $\Omega(\delta)_{D}$ (Eq.~\eqref{constrainedOmega}) for $^{18,22,26}$O. At a given value of the inverse temperature $\beta$, pairing correlations persist when $\delta$ is not too small. Pairing correlations keep the entropy low for moderate values of $\delta$ and even vanishingly small if the temperature is sufficiently low. However, as the zero-pairing limit is approached, the pairing energy carries less and less weight in the Routhian and the entropy plays a comparatively more important role. At high temperatures, this exchange of pairing energy for entropy is rather smooth, but for lower temperatures the process becomes more and more abrupt, i.e. pairing correlations need to almost collapse completely before the entropy term can dominate the grand potential. One observes that the limit value reached by the entropy at $\delta = 0$ equates the predicted value (Eq.~\eqref{entropylimits}) as $T \rightarrow 0$. While this limit value is zero in $^{22}$O, it is not in open-shell systems due to the non-zero contribution of the valence shell. 

Contrarily, when setting $T=0$ before taking the zero-pairing limit, the entropy vanishes in all nuclei whereas the pairing energy survives in open-shell systems as $\delta \rightarrow 0$ due to the residual anomalous density matrix within the valence shell. Once again, one observes the qualitatively different description when the order with which both limits are performed is inverted.

\section{Conclusions}
\label{conclusions}

The combined zero-pairing and zero-temperature limits of the finite-temperature Hartree-Fock-Bogoliubov formalism have been worked out analytically and realized numerically. The study is realized while imposing at least one, i.e. time-reversal, symmetry in order to be in the position to reach the non-trivial solutions of present interest\footnote{In a fully symmetry-fully-unrestricted calculation, all systems are necessarily of either spherical or deformed closed-shell character. For such  configurations, the present analysis leads only to trivial, i.e. textbook, solutions in the joint zero-pairing and zero-temperature limits.}. The present work extends the analysis of  Ref.~\cite{Duguet:2020hdm} where the zero-pairing limit of straight HFB theory was scrutinized.   

The textbook expectation is recovered for all closed-shell nuclei whether deformed or spherical: the FTHFB density operator reduces to a density operator corresponding to pure HF Slater determinant. For open-shell systems however, a non-trivial description is obtained. Furthermore, this non-trivial description is shown to depend on the order with which both limits are taken, i.e. the zero-pairing and zero-temperature limits do not commute in open-shell systems. When the zero-temperature limit is performed first, the FTHFB density operator is demoted to a projector onto a pure state, which is a linear combination of a finite number of Slater determinants with different (even) numbers of particles. When the zero-pairing limit is performed first, the FTHFB density operator remains a statistical mixture of a finite number of Slater determinants with both even and odd particle numbers. While the entropy (pairing density) is zero in the first (second) case, it does not vanish in the second (first) case in spite of being in the zero-temperature (zero-pairing) limit. To exemplify the consequences of the different descriptions obtained in the joint limits on the expectation value of operators, the particle-number variance was studied, i.e. it was shown to be different from zero and to differ by a factor of two in both cases. 

In summary, this analysis demonstrates that the behavior of the well-documented FTHFB formalism must be treated with care when taking its zero temperature and zero-pairing limits. In particular, this formalism, as opposed to the textbook expectation, does not reduce to Hartree-Fock theory {\it in all cases} when taking these joint limits. 

\section{Acknowledgments}
The authors warmly thank B. Bally, M. Bender, V. Som\`a and A. Tichai for their careful proofreading of the manuscript. W.R. gratefully acknowledges support by the U.S. DOE grant No. DE-SC0019521.

\begin{appendix}

\section{Useful formulae}
\label{formulae}

Newton's binomial formula along with its first and second derivatives with respect to $x$ provide three useful identities
\begin{align}
(x+y)^n &= \sum_{k=0}^{n} \binom{k}{n} \, x^{k} y^{n-k} \, , \label{binomial1} \\
n(x+y)^{n-1} &= \sum_{k=1}^{n} \binom{k}{n} \, k \, x^{k-1} y^{n-k} \, , \label{binomial2} \\
n(n-1)(x+y)^{n-2} &= \sum_{k=2}^{n} \binom{k}{n} \, k(k-1) x^{k-2} y^{n-k} \, . \label{binomial3}
\end{align}

\end{appendix}

\bibliographystyle{apsrev4-2}
\bibliography{zero_pairing}

\end{document}